\documentclass[11pt,letterpaper,parskip,bibtotoc]{article}

\usepackage[margin=1in]{geometry} 

\usepackage{setspace}
\usepackage{graphicx}
\usepackage{makecell}
\usepackage{csquotes}
\usepackage{float}
\usepackage{hyperref}
\usepackage{lineno}

\title{Cheating in online gaming spreads through observation and victimization}

\author{Ji Eun Kim and Milena Tsvetkova}

\date{}

\begin{document}

\maketitle

\vspace{-0.3cm}
\begin{center}
Department of Methodology, London School of Economics and Political Science, London WC2A 2AE, United Kingdom
\end{center}

\vspace{0.5cm}

\section*{Abstract}

Antisocial behavior can be contagious, spreading from individual to individual and rippling through social networks. Moreover, it can spread not only through third-party influence from observation, just like innovations or individual behavior do, but also through direct experience, via ``pay-it-forward'' retaliation. Here, we distinguish between the effects of observation and victimization for the contagion of antisocial behavior by analyzing large-scale digital-trace data. We study the spread of cheating in more than a million matches of an online multiplayer first-person shooter game, in which up to 100 players compete individually or in teams against strangers. We identify event sequences in which a player who observes or is killed by a certain number of cheaters starts cheating, and evaluate the extent to which these sequences would appear if we preserve the team and interaction structure but assume alternative gameplay scenarios. The results reveal that social contagion is only likely to exist for those who both observe and experience cheating, suggesting that third-party influence and ``pay-it-forward'' reciprocity interact positively. In addition, the effect is present only for those who both observe and experience more than once, suggesting that cheating is more likely to spread after repeated or multi-source exposure. Approaching online games as models of social systems, we use the findings to discuss strategies for targeted interventions to stem the spread of cheating and antisocial behavior more generally in online communities, schools, organizations, and sports.

\vspace{0.5cm}
\noindent\textbf{Keywords:} social contagion, antisocial behavior, temporal networks, online multiplayer games 

\vspace{1cm}

\section*{Introduction}

We, humans, are largely a product of our natural and social environment. Our bodies consist of at least as many foreign bacterial cells as our own \cite{sender16}. Similarly, our emotions \cite{kramer14}, political decisions \cite{gonzalez11,bond12}, health behavior \cite{centola10,aral17}, knowledge \cite{muchnik13,vosoughi18}, and even memories \cite{patihis13} are influenced by our friends and peers. Others influence not only our individual behavior but also our social behavior, i.e., how we behave towards other people. 

Previous research has shown that prosocial, as well as antisocial behavior can spread from one individual to another, even among strangers. Prosocial behavior confers a benefit to others that comes at a personal cost, while antisocial behavior imposes costs on others often for one's own personal gain. Prosocial behavior is contagious, such that if you help a stranger, you not only increase the likelihood that they help others \cite{tsvetkova14,kizilcec18,norbutas18,simpson18} but possibly those they help will also help others, and so on, out to three steps \cite{fowler10}. But unfortunately, socially irresponsible behavior such as littering, stealing, verbal aggression, and cheating spreads in a similar manner \cite{cialdini90,falk02,keizer08,gino09,rettinger09,cheng17}.  

Most of the prior research investigates the contagion of antisocial behavior through observation. The idea is that when witnessing antisocial behavior by others, the observer may change their estimation of the likelihood of being caught or change their understanding of the social norms related to that behavior \cite{cialdini90,gino09}. For instance, the ``broken windows'' hypothesis posits that this is how minor infractions such as graffiti, litter, or vandalism multiply and escalate in neighborhoods \cite{wilson82}. However, antisocial behavior may also spread through victimization \cite{tsvetkova15}. In this case, those who are victims of antisocial behavior may ``pay it forward'' by ``retaliating'' not against the perpetrator but against a third party. Driven by negative affect, such as feelings of injustice, anger, and frustration, victims may displace their aggression and involve innocent others \cite{hoobler06}.  

This study extends research on the contagion of antisocial behavior by empirically distinguishing between the effects of observation and victimization with observational data. We investigate the spread of cheating in an online multiplayer first-person shooter game and hypothesize that players who observe or are killed by a cheater are more likely to start cheating, particularly if they both observe and experience cheating. We further hypothesize that the effects of observation, victimization, and their positive interaction will be stronger with repeated exposure. To overcome the challenges of studying social influence in real social networks, we employ novel computational analysis techniques on a unique large dataset of gameplay logs. For the analyses, we count all possible event sequences indicative of contagion and test whether we can explain them away with alternative gameplay scenarios while accounting for homophily, unobserved individual characteristics, and unobserved external factors.   

Our research demonstrates the power and promise of computational methods to identify social mechanisms and causal processes in large-scale network data. Our empirical findings illuminate important and fundamental social processes that manifest themselves in many different social contexts. If we perceive online games and virtual worlds as a simulation or abstraction of real-world social and economic transactions \cite{bainbridge07,szell10}, our findings relate to antisocial behavior such as negative gossip in the office, academic cheating, bullying, illegal drug use by athletes, and employee theft. The contagion of antisocial behavior implies that a single act of misbehavior has the potential to trigger a chain reaction that reaches far beyond the original initiator. To prevent this from happening, we need to understand when to intervene, whom to target, and how. This knowledge will aid policy makers, managers, administrators, and educators develop effective strategies to reduce the incidence and normative acceptance of antisocial behavior and ensure functional and sustainable organizations and communities.

\subsection*{Mechanisms for the contagion of social behavior}

There is an ever-growing literature on social influence and contagion, which investigates the spread of opinions and behaviors over social networks \cite{rogers03,christakis09,sinclair12}. Here, we narrow our attention to a specific subarea of the problem: the contagion of social behavior among strangers. In contrast to individual behavior, social behavior relates to social interactions. Thus, social behavior can spread not only through observation, similarly to when adopting an innovation or a healthy habit, but also through direct experience, when being subjected to that very behavior. These two separate contagion pathways correspond to two distinct behavioral mechanisms identified by previous research: third-party influence and generalized reciprocity \cite{tsvetkova14,tsvetkova15}. 

Third-party influence refers to cases in which those who observe certain social behavior emulate it. This mechanism characterizes social learning via imitating others \cite{cialdini90,gino09} and corresponds to informational influence \cite{deutsch55}. Generalized reciprocity refers to cases in which those who directly experience certain social behavior adopt it. Generalized reciprocity characterizes a displaced response triggered by normative obligation or affect---positive affect and gratitude in the case of prosocial behavior \cite{bartlett06}, and anger, frustration, hostility, aggression, and desire to revenge in the case of antisocial behavior  \cite{hoobler06}. Generalized reciprocity is better described as normative influence---an urge to act as the other did and, in the lack of opportunities to repay the act, to let it out, to pay it forward. 

Generalized reciprocity, as we use the term here, is related to what the literature on cooperation refers to as ``upstream indirect reciprocity'', as opposed to ``downstream indirect reciprocity'' \cite{nowak05,nowak07}. In both cases, we expect a correlation between what one experiences and how one behaves towards others but the motivations are entirely different. Upstream indirect reciprocity is reactive---a normative or affective response to a recent experience, while downstream indirect reciprocity is proactive---an instrumentally motivated strategy regarding one's reputation or ``image-score''  in anticipation of future interactions \cite{wedekind00,bartlett06,nowak05}. A behavior could increase with downstream indirect reciprocity because reputation modifies the incentives in the interaction situation. However, it is upstream indirect reciprocity that causes social contagion, in the sense of transmission upon contact. Yet, in practice, the two mechanisms are often confounded. Since most social systems involve repeated interactions in fixed network structures and with knowledge about others' past behavior, they facilitate reputational motivations. For example, in generalized exchange systems, such as the Kula trading ring among South Pacific islanders \cite{malinowski20} and the kinship relations among aboriginal tribes \cite{bearman97}, the observed chains of helping behavior are likely driven by anticipated rewards, rather than an affective response to receiving help. Generalized reciprocity is more likely when interactions are anonymous and transient and reputational information unavailable or non-salient \cite{ben-ner04,stanca09,tsvetkova14,tsvetkova15}.

Third-party influence and generalized reciprocity are conceptually distinct but they are likely to interact, due to the greater likelihood to both observe and personally experience behavior that is common in the population. In fact, to experience social behavior also means to observe it, although one can observe social behavior without experiencing it. Thus, in practice, generalized reciprocity cannot be completely isolated from third-party influence. Nevertheless, if we compare the response to experiencing, observing, and both experiencing and observing, we can provide evidence for either mechanism, as well as their interaction. Experiencing and observing are expected to independently increase the adoption of certain social behavior, but because they activate different mechanisms, their combination could have even stronger effect.   

We further expect the two mechanisms, as well as their interaction, to be stronger when one observes or experiences on multiple occasions. There could be two reasons for this. First, if we assume that social behavior spreads upon contact like simple contagion, the probability to adopt increases with repeated contact. For example, just as you are more likely to contract a contagious disease from a household member than a business client, you are more likely to mimic the social attitudes and behaviors of someone you encounter repeatedly compared to those with whom you interact only occasionally. Second, if we assume that social behavior spreads like complex contagion, adoption will be more likely after exposure to multiple adopters \cite{centola07,centola18}. Individuals are more likely to adopt costly and risky behaviors if they know that they are common and acceptable, and social behavior is costly and risky: prosocial behavior such as rendering help or choosing to cooperate requires resources, effort, time, and attention, while antisocial behavior such as cheating, stealing, or shirking contributions risks future sanctions and punishment. Although, in the context of anonymous interactions among strangers, it is impossible to differentiate between repeated exposure to one adopter and exposure to multiple adopters, both processes point to the same overall effect: adoption is more likely after repeated observation and/or victimization compared to a single exposure.

\subsection*{The contagion of antisocial behavior}

Most of the research on the contagion of antisocial behavior overlooks the distinction between third-party influence and generalized reciprocity. As a result, existing empirical studies either conflate the effects of observing antisocial behavior with becoming a victim of antisocial behavior, or provide evidence for contagion via observation only.

Criminologists and scholars of deviance were among the first to argue that antisocial behavior can be contagious. In alignment with the argument here, they hypothesized that violent behavior can be transmitted because people learn it socially through communication and interaction \cite{sutherland92} and those who witness or experience violence retaliate or pre-empt it \cite{loftin86}. Consistent with these hypotheses, observational studies reveal that violent crime is spatially clustered, tends to be reciprocated at the individual level, and tends to escalate at the group level \cite{loftin86}. However, the observed correlations cannot rule out alternative mechanisms such as a common external cause or contextual factors \cite{fagan07} and overall, the empirical support for the interpersonal transmission of violent crime is contradictory and problematic \cite{widom89}.

The empirical support for the contagion of antisocial behavior is much more robust and consistent in the context of non-violent crime. Much of the evidence comes from tests of the ``broken windows'' hypothesis \cite{wilson82}. The theory posits that minor infractions such as graffiti, litter, or vandalism can signal the absence of monitoring, enforcement, and public support of laws and social norms, encouraging such behavior, and leading to a self-reinforcing downward spiral. Field experiments have revealed that if people witness minor norm and law violations, they become more likely to litter, trespass, and steal \cite{cialdini90,keizer08}. In the lab, participants who observe other group members steal, cheat, or defect in cooperation games become more likely to do that too \cite{falk02,gino09,jordan13,dimant19}. In schools, the presence and awareness of cheaters are associated with higher rates of academic cheating \cite{mccabe01,carrell08,rettinger09}. And in online forums, incidents of trolling and swearing cause others to exhibit verbal aggression \cite{cheng17,kwon17}. 

Only one previous experimental study of antisocial behavior distinguishes between third-party influence and generalized reciprocity \cite{tsvetkova15}. Participants in the experiment are given the opportunity to transfer resources from others, where some participants have lost resources in this way, others have knowledge of such cases of theft occurring, and still others are both victims and observers. The study finds support for generalized reciprocity but not for third-party influence, nor an interaction effect between the two. Further, the experiment does not test for effects from repeated victimization.   

Isolating generalized reciprocity from third-party influence is already challenging with an experimental setup but becomes particularly so with observational data. Real social networks present three potential confounders to social influence  \cite{aral09,shalizi11}. First, social interactions are rarely random but driven by a number of social mechanisms, one of which is homophily, or the tendency to associate with similar others. Thus, similarity in behavior between two individuals could be the reason for interaction between them, not the product of it. Second, interacting individuals are likely to be exposed to similar situational factors we cannot observe and these could be the ones actually responsible for temporally proximate changes in behavior. Third, unobserved individual characteristics or situational factors could be driving both the interaction and the propensity for the particular behavior, creating a spurious relationship between the two. 

Even if we can identify social influence, the challenge to isolate the precise mechanisms remains. First, observation and victimization are often confounded: for example, they are both driven by the individual's social activity and the level of antisocial behavior in the individual's immediate social environment. Second, the effect from victimization may not only capture generalized reciprocity but also direct reciprocity (revenge) and reputation-based indirect reciprocity (e.g. I exhibit violent behavior to deter others from attacking me).  

This study uses advanced computational methods on large-scale digital trace-data from a unique social setting to overcome these challenges. We investigate the spread of cheating in an online multiplayer game and look for empirical evidence that players who observe cheating, experience cheating, both observe and experience, and observe and/or experience multiple times become more likely to cheat.

\subsection*{The contagion of cheating in online gaming}

In online gaming, cheating involves the use of unauthorized software or hardware to modify certain in-game elements. Gamers use cheating tools to enhance their abilities and increase their chances of winning, thus harming the game performance and experience of other players. These tools can be purchased from online providers or downloaded for free. Many different types of cheating tools exist: for example, ``aimbots'' automatically aim a gun at other players and ``wallhacks'' make walls transparent to allow players to see or attack competitors hiding behind a wall. Cheating in online gaming is analogous to using performance-enhancing drugs in professional sports or cheating on a test in school. Correspondingly, providing cheating code for online games constitutes a lucrative but illegal business. From the perspective of game companies, cheating presents serious problems, including damaged reputations and loss of revenue. Many game companies develop anti-cheating solutions to detect cheaters and punish them by flagging \cite{blackburn14,zuo16} and even suspending or permanently banning \cite{kakao20} their accounts. Thus, the war between game companies, cheat code providers, and cheating players is ongoing.

Most online game players are not professionals and play for fun, to socialize with friends, or to boast skills and performance \cite{yee06}. This raises the question of why cheating occurs at all. Interviews, focus groups, and surveys suggest that players who cheat disregard cheating as a serious problem, want to annoy other players, want to progress in the game, or desire to create a new gaming experience \cite{consalvo09,wu13}. These studies, however, also find that exposure to friends who cheat is closely related to adopting cheating \cite{wu13,chen18}. In other words, cheating may be socially contagious. Supporting this hypothesis, Blackburn and colleagues \cite{blackburn14} use data from Steam, a digital game distribution platform where players can purchase games and make friends, and find that the number of cheating friends predicts the future adoption of cheating. Similarly, Zuo et al. \cite{zuo16} report that non-cheaters are more likely to adopt cheating if they are exposed to unpunished cheating neighbors in the Steam Community or to cheating teammates in the game. 

These studies support the hypothesis that players who have many cheating friends may start cheating as well but do not distinguish the effects of social influence from homophily. Woo et al. \cite{woo18} attempt to do this by comparing the correlation between number of cheating friends and cheating adoption in the observed network with the one observed when adoption times are shuffled. This method confirms that cheating is socially contagious among friends in the massively multiplayer online role playing game they study. An even more promising approach to remove homophily is to study influence among opponents, rather than friends, since many games match opponents randomly. Zuo et al. \cite{zuo16} do this and show that players who are randomly matched with more cheaters are more likely to adopt cheating. However, the authors do not account for the fact that players who play more and are hence more likely to interact with cheaters may be also more likely to adopt cheating for entirely different reasons. Our research addresses this caveat.

We study cheating in PlayerUnknown's Battlegrounds (PUBG), a popular online multiplayer first-person shooter game (Fig.\ S1). First-person shooter is a game genre in which players use weapons to kill other players' characters in the first-person point of view. A typical PUBG match involves up to 100 players. At the beginning of the match, the players start from different spots on the virtual battlefield. They scavenge the field for weapons and kill other players they encounter to survive and win. Players can make a car, motorcycle or boat trip to move from one place to another. To prevent players from deliberately avoiding combat or hiding for a long time, the game constantly reduces the size of the battlefield over time. If players do not move to the safe area, they are harmed by a player-damaging barrier or bombs and eventually get killed. Thus, players ultimately end up in one place at the game's conclusion after moving constantly to avoid danger zones. The last survivor becomes the winner of the match.

PUBG randomly assigns players to a match and players cannot choose their opponents. However, when the match is played in a team mode, players can invite friends to form a team of up to two (duo matches) or four (squad matches) (Table S1); if the player does not select any friends, they could be assigned random teammates. Matches are independent in the sense that players' characters do not carry over achievements from one match to another and everyone starts the match in the same state. Performing well in a match requires the player to plan strategies, move their character, and react to other players' actions in real time. Generally, players improve their performance by playing repeatedly and learning from experience; some players also watch professional PUBG games or YouTube videos of good players in order to learn skills and strategies faster. PUBG maintains a skill-based ranking system for players, one for each of the three game modes: solo, duo, or squad. The ranking method is complex but overall, rank improves when the player places higher in a match, and to a lower extent, when the player achieves more kills.

When it comes to cheating, PUBG is similar to other popular online first-person shooter games: the use of cheating tools is common, problematic, and penalized. The game, however, offers a unique opportunity to study social contagion via observation and victimization because it precludes or limits other possible mechanisms for the adoption of cheating. Specifically, PUBG does not facilitate direct reciprocity, nor indirect reputation-based reciprocity. Players are randomly matched to opponents, out of tens of thousands of players who play per day on a particular game server, and thus unlikely to encounter each other again. For this reason, victims are unlikely to adopt cheating in order to take revenge on their cheating killer. Similarly, players are unlikely to adopt cheating in order to build a reputation as someone who punishes cheaters: such reputations cannot form and persist between matches.   

Nevertheless, due to the team structure, similarity in cheating adoption could be driven by homophily rather than third-party influence. Moreover, even though cheaters are randomly assigned to matches, the exposure to a cheating opponent is not entirely random. Players who play more often or even who play at certain times of the day or days of the week may be more likely to be matched with a cheater. Further, becoming a victim or observing a cheater are not independent treatments since, in a match, players can get killed once only and the killed can neither kill nor observe. To provide a strict test for the social contagion of cheating via observation and victimization, our analyses account for the temporal and structural patterns of the interactions within matches and across the game.

\subsection*{Hypotheses}

To summarize, we study the adoption of cheating in the online game PUBG in order to investigate whether antisocial behavior spreads among strangers via observation and victimization. On the one hand, the well-established idea of third-party influence, or imitation, suggests that antisocial behavior could spread when those who observe such behavior emulate it. Specifically, online game players who observe a cheater might notice the power of cheating tools and possibly conclude that cheating is acceptable and unpunished, and as a result become more likely to start cheating. Hence, we first test the following hypothesis: 

\emph{H1. Observing a cheater increases the likelihood to start cheating.}

On the other hand, the idea of generalized reciprocity, uniquely applicable to the contagion of other-directed behavior, suggests that antisocial behavior spreads when its victims pay it forward. In an online game such as PUBG, players who are killed by cheaters might experience anger or frustration that could make them more likely to harm other players by cheating. We thus hypothesize that:

\emph{H2. Being killed by a cheater increases the likelihood to start cheating.}

Since third-party influence and generalized reciprocity are distinct contagion mechanisms, they might reinforce each other and produce an effect that neither can instigate alone. Consequently, we hypothesize that, regardless of whether observers or victims of cheating start cheating, players who have both observed and being killed by a cheater are more likely to start cheating: 

\emph{H3. Both observing and being killed by a cheater increases the likelihood to start cheating.}

Finally, we hypothesize that repeated exposure increases the likelihood of adoption. Since we focus on anonymous interactions, we cannot distinguish between repeated exposure to the same adopter and one-time exposures to multiple adopters. Nevertheless, in either case---repeated simple contagion or complex contagion---repeated exposure is expected to facilitate the spread of cheating. In parallel to H1--H3, we formulate the hypothesis separately for players who have repeatedly observed, repeatedly suffered from, and both observed and suffered from cheaters:

\emph{H4.1. Observing a cheater on more than one occasion increases the likelihood to start cheating.}

\emph{H4.2. Being killed by a cheater on more than one occasion increases the likelihood to start cheating.}

\emph{H4.3. Both observing and being killed by a cheater on more than one occasion increases the likelihood to start cheating.}

We note that the hypotheses are formulated to test against the null hypothesis of absence of effect but do not address the relative size of effects. This limitation is dictated by the analytical approach we employ, as described next. 

\section*{Data and Methods}

The study was reviewed and approved by the London School of Economics and Political Science. 
Aggregate data and all scripts used for the analyses can be accessed on GitHub:  \url{https://github.com/social-research/cheating}. 
The complete anonymized individual-level data is also available \cite{kim21}. All analyses were carried out using Spark, Python, and R.

Our data come from two different online sources. First, in compliance with their terms of use, we accessed the PUBG API to obtain gameplay logs from the Kakao Games server, the smaller of the two gaming platforms that distribute PUBG in South Korea (the other being Steam). Second, we downloaded the lists of banned cheaters that Kakao Games uploads daily on its public website.  

We collected data on all 1,291,441 matches that were played on the server between March 1 and March 31, 2019. Each match is described with a link to a telemetry file, which contains a list of events that happened during the match and detailed logs. Due to technical issues with the API, it was impossible to decode about 9\% of the telemetry files. We further removed 32,596 matches that were played in special modes such that players could revive multiple times. As a result, the final dataset we analyze consists of 1,146,941 matches.

\subsection*{Measures and operationalizations}

\subsubsection*{Cheating}

To identify cheaters, we rely on how Kakao Games defines cheating and how it identifies and reports cheaters. In its rules of conduct, the game company states that cheating is using ``unauthorized programs and hardware devices that are not permitted or authorized by the company'' on its website \cite{kakao20}. Unfortunately, the company does not list the specific cheating methods or tools it prohibits. The game is monitored around the clock by an anti-cheat system which automatically checks for abnormalities caused by cheating. Also, the company solicits reports from players who have encountered cheaters through its in-game reporting system (Fig.\ S1-B). If Kakao Games finds evidence of cheating in the data reported, it bans the cheater. Players who are banned are no longer able to purchase another copy of the game. The company is sensitive to false positives as it may lose customers when banning innocent players. Thus, players who were banned almost certainly cheated in the game, although it is possible that many other cheaters were missed.

\subsubsection*{Estimating the time of cheating adoption}

The game company provides the date when a ban is applied to a cheater but not when the player started cheating. To estimate when a banned cheater started cheating, we analyze the players' behavior. In general, cheaters tend to show abnormal and noticeable kill patterns \cite{alayed13}. We thus focus on two features related to kill scores that have been previously used for cheating detection: the average kill ratio per day and the average time difference between consecutive kills per day. The average kill ratio per day is calculated by dividing the number of kills by the sum of the number of kills and deaths in a day. The average time difference between consecutive kills is measured if a player killed at least two other players during the match.  

To confirm that cheaters and non-cheaters are fundamentally different in terms of performance, we compare the two groups by tentatively supposing that cheaters who were banned between March 1 and March 3, 2019 were cheating during those three days. The number of cheaters who were identified and banned during this period is 651, while the number of non-cheaters who played at least one match then is 854,153. As expected, we find that cheaters show higher average kill ratios than other players (Fig.\ S2-A). The mean values of cheaters and non-cheaters are 0.77 and 0.40 and the median---$0.82$ and 0.44, respectively. Using Welch's two-independent-samples t-test to compare the two groups, we obtain $t(651.26$ df$) = -48.64$, $p < 0.01$, which is statistically significant. Further, as expected, the average time difference between kills is shorter for cheaters than non-cheaters (Fig.\ S2-B). The mean values for cheaters and non-cheaters are 139.67 and 194.11 seconds, respectively and the median---$123.93$ and 172.63. The difference between the two is statistically significant: Welch's $t(632.74$ df$) = -18.24$, $p < 0.01$. 

On the basis of these observations, we develop a rule-based algorithm to estimate the starting date of cheating. We assume that cheaters start cheating on the day when they meet the following two conditions: 1) average kill ratio $\geq$ 0.8 and 2) average time difference between consecutive kills $\leq$ 140 seconds. Among the 6,161 identified cheaters, we have complete performance information for 2,980. For these, the average duration of cheating before the ban is four days, with a modal value of two days (Fig.\ S3). For the remaining 3,181 cheaters who have at least one missing value on performance or do not meet the conditions above, we apply the modal value of two days as the period of cheating. As a result of this approximation, around 70\% of all cheaters (4,321 players) were estimated to have started cheating two days prior to being banned.

\subsubsection*{Recognizing cheaters}

Our definition of observation and victimization relies on the ability of players to identify who is a cheater but this is not guaranteed since cheaters are not flagged in the game. Cheaters often exhibit abnormal kill patterns which can give them away. For example, players would recognize a cheater when they are killed by a bullet that penetrates a wall or when they observe another player move abnormally fast, dispose of unlimited ammunition, or kill many other players consecutively in a very short time interval. In PUBG, all players can see who killed whom in real time through ``kill feeds'' that appear on the upper right corner of the screen during the match (Fig.\ S1-A). In addition, PUBG offers an in-game system that allows players to watch replay recordings or watch the rest of the game from the perspective of their killers, and the game company actively encourages players to use this system and report cheaters. 

To affirm the robustness of the effects we study to measurement errors, we use two different definitions of observation and victimization. In the simple, more relaxed definition, we assume that observation occurs when the cheater has killed at least two other players before the observer is killed, while victimization occurs any time a player is killed by a cheater. 

In the strict definition, we assume that only players who die after a cheater has killed at least five other players observe the cheater, while only players among the 30\% remaining in the game and killed by a cheater are victimized. Thus, the strict definition posits that a player would identify a cheater only if the levels of observation and harm are salient enough for them to pay attention and investigate. Alternatively, the strict definition implies that social contagion only occurs when the stimuli are sufficiently strong. Put differently, the strict definition provides more reliable measurements of the phenomena, but also restricts the analysis to more impactful events.

\subsection*{Analytical approach}

\begin{figure}[h]
\centering
\includegraphics[width=\textwidth]{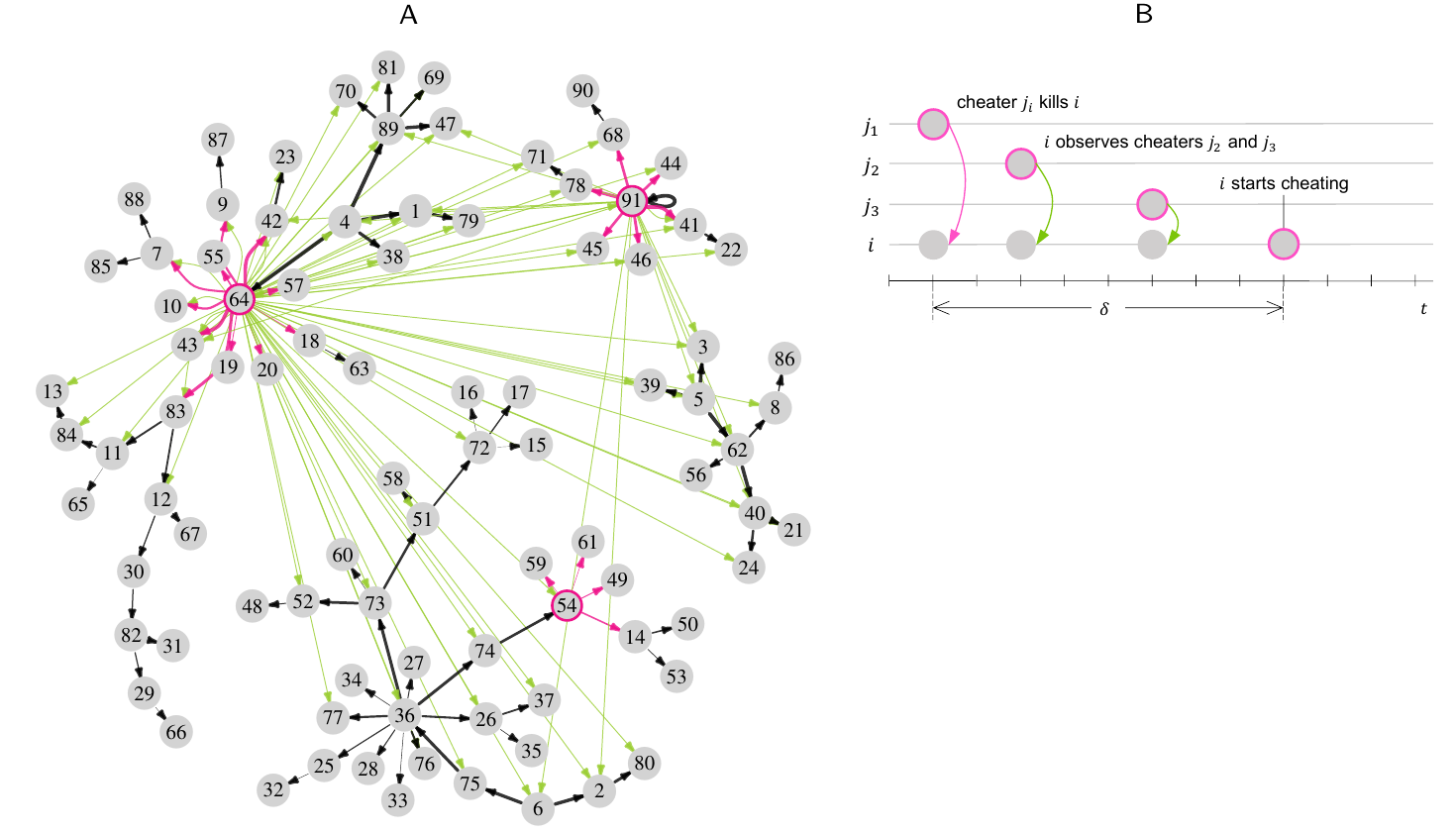}
\caption{A) PUBG matches can be seen as temporal networks in which an edge points from the killer to the victim (black or pink) and from the cheater to the observer (yellow-green). For clarity, the visualization shows that an observation occurred when the cheater killed at least five times while the player was in the game (this corresponds to the strict definition of observation we use in this study). Cheaters and their killings are outlined in pink. Thicker edges show killing events that occur later in the match. The winner of this example squad match is the team composed of nodes 4, 5, and 6. The self-loop for player 91 indicates that they either killed themselves or died due to a player-damaging barrier or bombs in danger zones. B) We identify event sequences in the killing networks in which player $i$ is killed by and/or observes cheaters before $i$ starts cheating within a period $\delta = 7$ days.}
\label{fig:fig1}
\end{figure}

The gaming interactions constitute relational events: discrete timestamped actions directed from one player to another \cite{butts08}. We distinguish between two types of relational events: killing a player or observing a cheater. Thus, we also assume that players have a timestamped attribute that determines whether they are a cheater or not and we identify non-relational events whereby a non-cheating player becomes a cheater. The sequence and interconnection of relational events in each match can be represented as a temporal network \cite{holme12} (Fig.\ \ref{fig:fig1}-A).

\subsubsection*{Contagion event sequences}

The characteristic signature of the social contagion of cheating is a sequence of events in which a non-cheating player observes $n_o$ cheaters and/or is killed by $n_v$ cheaters, with at least one of $n_o$ and $n_v$ greater than zero, and then becomes a cheater within some time period $\delta$ (Fig.\ \ref{fig:fig1}-B). This sequence denotes not only the fact that the cause (observing and/or experiencing cheating) should precede the effect (adopting cheating) but also the idea that the two should be temporally proximate. Given the typical weekly cycle of gameplay activity in our data (Fig.\ S4-A), we assume a time window for influence $\delta =7$ days. Further, the $n_o$ and $n_v$ sets could partially overlap since, within a match, it is possible to observe a cheater before being killed by them.

Event sequences of this type, however, can occur merely due to chance or due to homophily and unobserved individual and situational factors that make interactions among cheaters likely. If this were the case, any network with equivalent player composition and topology would yield a similar number of such event sequences. We can thus use suitably randomized versions of the empirical network to estimate the expected count of contagion event sequences under these baseline conditions. The null hypothesis is that the number of contagion event sequences in the empirical network is not larger than the one expected in a randomized network. Rejecting the null would suggest that at least some of the observed event sequences result from a different social process, and we posit that this process is social contagion. 

We note that although we analyze relational events, unlike relational event models \cite{butts08}, we do not aim to explain their structure but to test if they predict cheating-adoption events. Our approach is more closely related to temporal motif analysis. Network motifs are subgraph patterns with a fixed topology that occur more frequently in the real network than in random networks \cite{milo02}. In temporal networks, temporal motifs are topologically equivalent patterns that additionally include the same order of events, confined to occur within a specified period of time \cite{kovanen11,paranjape17}. If we assign colors to nodes \cite{kovanen13,ribeiro14}, then we can represent social contagion as motifs in which a node changes color after interacting with a colored node. Although the event sequences we study here represent elementary subgraph structures, temporal motif analysis allows to extend our research to explore longer and more complex chains of influence.

To test the independent positive effect from observation on the adoption of cheating (H1), we estimate whether contagion event sequences with $n_o > 0$ and $n_v = 0$ occur significantly more often in the empirical data than in the randomized networks. For the expected independent effect from victimization (H2), we do the same for sequences with $n_o = 0$ and $n_v > 0$. To test the hypothesis on the positive interaction between observation and victimization (H3), we estimate whether sequences with both $n_o > 0$ and $n_v > 0$ are overrepresented compared to the null model. Finally, we test for the effects of repeated exposure by confirming that sequences with $n_o \geq 2$ and $n_v = 0$ (H4.1), $n_o = 0$ and $n_v \geq 2$ (H4.2), and $n_o, n_v \geq 1$ with either $n_o \geq 2$ or $n_v \geq 2$ (H4.3) are more overrepresented and less likely to occur in the null model.

\begin{figure}[h]
\centering
\includegraphics[width=\textwidth]{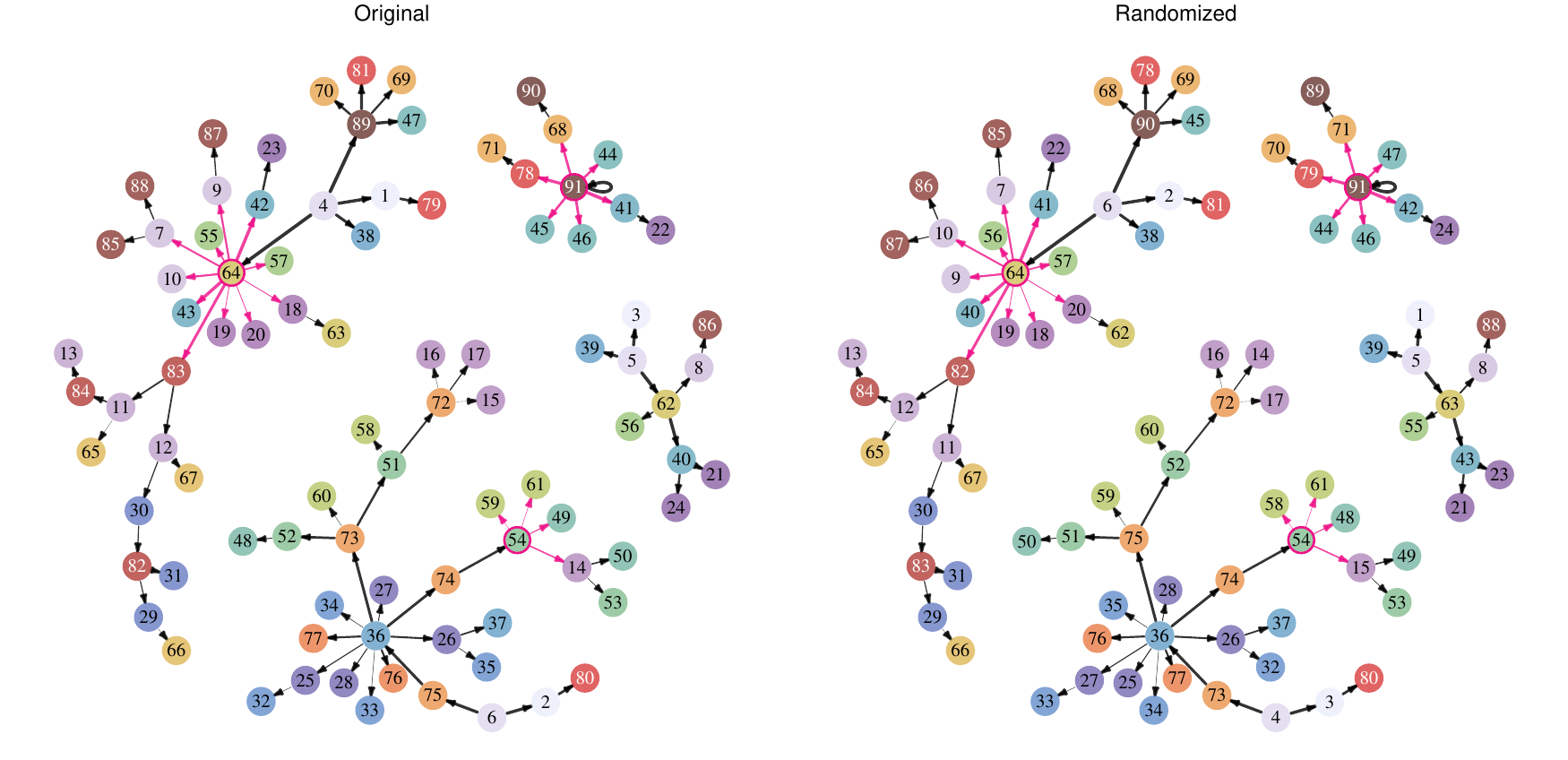}
\caption{To simulate alternative sequence of events, we permute the nodes in the killing networks by match. Permutation is constrained within cheater-type and within teams to account for the fact that cheaters tend to kill more and that team members tend to move together and avoid killing each other. In this example of permuting a squad match, nodes with the same color belong to the same team. Cheaters (nodes 54, 64, and 91) and their killings are outlined in pink.}
\label{fig:fig2}
\end{figure}

\subsubsection*{Null model}

We use a simple and conservative null model that preserves aspects of the interactions that could indicate alternative mechanisms for the adoption of cheating but simulate alternative interaction sequences to use as counterfactuals to the hypothesized causal process for social contagion (Fig.\ \ref{fig:fig2}). The null model uses node-label permutation \cite{croft11}: for each match, we preserve the event structure but reshuffle the player ids, with certain constraints. We permute cheaters in a match separately from non-cheaters in order to account for the fact that cheaters kill more opponents than non-cheaters and thus exhibit higher node outdegrees (Fig.\ S2-A). Since 87.3\% of the 107,139 matches with at least one cheater have \textit{exactly} one cheater, in most cases, cheaters remain in the same position as in the original network. In addition, if the match is played in team mode, we permute players within teams only (Fig.\ \ref{fig:fig2}). This accounts for social influence or homophily from friends since cheaters are unlikely to kill their teammates but more likely to be observed by them as teams typically move together. Essentially, we randomly reassign who the victims and observers are in any particular match, subject to the constraints that team members stick together and cheaters tend to kill more. 

We repeat the randomization process $n_r = 1000$ times and count the contagion event sequences in each instance. Previous research typically uses $z$-scores to evaluate the deviation of the empirical counts from the null model \cite{kovanen13,tsvetkova16}. However, since we investigate rare events (Fig.\ S5) and use a significantly constrained null model, neither the normal nor the Poisson distributions present suitable statistical approximations for the expected value of the counts (Fig.\ S6 and S7). Consequently, we directly estimate the proportion $\hat{p}$ of our sample of $n_r$ randomized networks that produce a count that is equal to or larger than the one in the empirical network. We additionally calculate the standard error for the proportion using the formula $\sqrt{\frac{\hat{p}(1 - \hat{p})}{n_r}}$. 
This gives us an estimation of the probability $p$ (and our certainty about it) to observe contagion event sequence counts as large as the empirical ones under the null hypothesis. We assume $p < 0.05$ is sufficient to reject the null hypothesis. Overall, we conduct eight tests for the hypotheses: one each for H1, H2, H3, H4.1, and H4.2 and three for H4.3, and then replicate these tests for two different operationalizations of observation and victimization. Since the eight tests are dictated by planned and theoretically informed hypotheses and the replications are complementary, we do not correct for multiple testing. Nevertheless, we discuss the consequences of such correction when we present the results for H4.3 and revisit the issue in the discussion.

\section*{Results}

Our data cover 1,146,941 matches played by 1,975,877 unique players and involving 98,319,451 killings. Only a small portion of the players played the game regularly every day; 30\% of the players accessed the game just on a single day, with a median participation period of three days (Fig.\ S4-B). Over our observation period, 6,161 of the players, equivalent to 0.3\%, were banned for cheating, 286,914 players were killed by a cheater at least once, and 1,185,279 players played in at least one game with a cheater. Cheaters represent a small proportion of all players but are responsible for disproportionately more killings: 0.46\% of the killings that were not self-kills were done by cheaters (453,071 out of 97,841,185; in comparison, cheaters were killed 98,349 times). Moreover, thousands of teams have two or more cheating members (Table S1), providing evidence for homophily and/or influence among friends regarding cheating. Further, since teams with more cheating members are more successful (Table S2), players likely aim to team up with cheaters. The null model we use preserves the composition of teams in order to account for these competing mechanisms and identify the contagion of cheating among strangers.

\begin{figure}[h]
\centering
\includegraphics[scale=0.71]{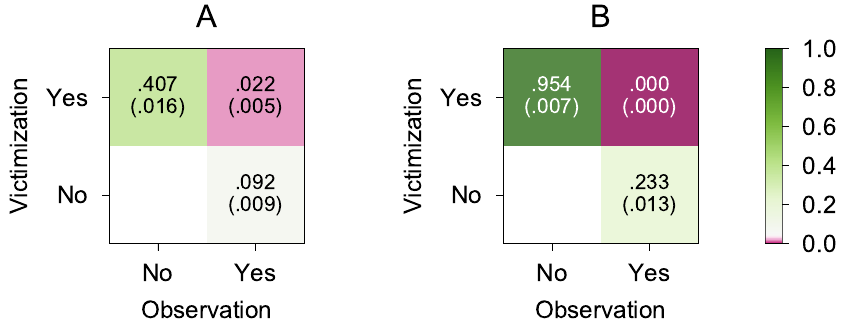}
\caption{Cheating is contagious only for those who both observe and suffer from it. Cell numbers and colors show the probability (with the standard error in brackets) to observe contagion event sequence counts as large as the empirical ones in a randomized network, estimated over 1,000 randomizations for A) simple and B) strict definitions of observation and victimization. Under the simple definition, observation occurs when a cheater kills at least two others while the player is still in the game and victimization occurs every time a player is killed by a cheater. Under the strict definition, observation occurs when a cheater kills at least five others while the player is still in the game and victimization occurs if a player is killed by a cheater when the player is among the last 30\% of survivors.}
\label{fig:fig3}
\end{figure}

First, we test for the effects of observation, victimization, and their interaction (H1-3) by counting contagion event sequences regardless of whether they involve the same nodes. For example, if a player was never killed by a cheater but observed two cheaters, we count this as two observation-only event sequences; if a player was killed by three cheaters and observed four cheaters, we count this as three observation-and-victimization event sequences. We find that the probabilities to observe event sequence counts as large as the ones we have in the empirical networks compared to what we would expect in networks with randomized events are sufficiently small only for the observation-and-victimization event sequences (Fig.\ \ref{fig:fig3}). The results are consistent for both the simple and strict definitions of observation and victimization. This suggests that although we do not find evidence for a direct effect from observation alone (H1), nor for a direct effect from victimization alone (H2), their co-presence appears to be associated with adopting cheating more than expected by chance, which supports the positive interaction effect hypothesized in H3.

\begin{figure}[h]
\centering
\includegraphics[scale=0.71]{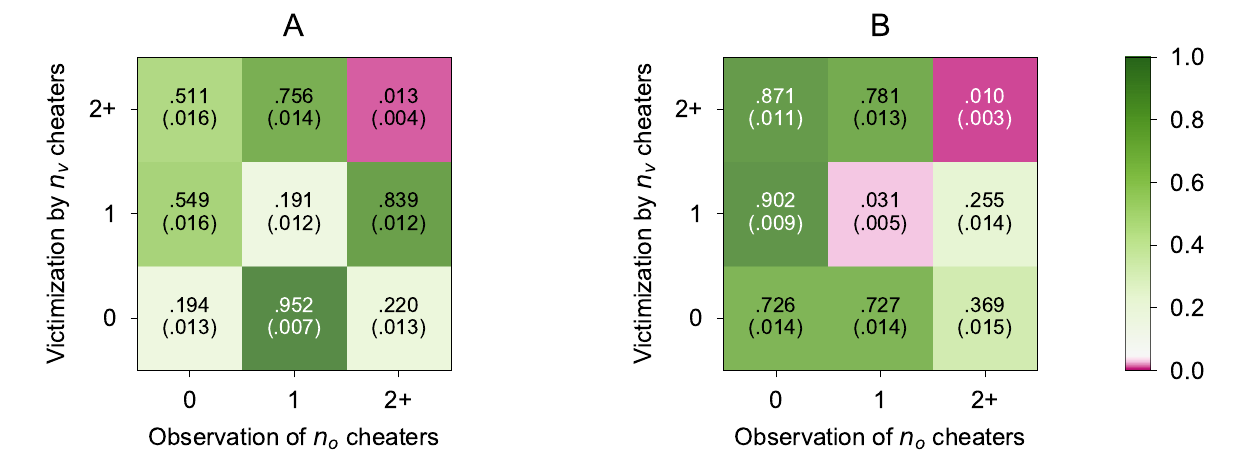}
\caption{Cheating is contagious for those who both observe and suffer from it on more than one occasion. Cell numbers and colors show the probability (with the standard error in brackets) to observe contagion event sequence counts as large as the empirical ones in a randomized network, estimated over 1,000 randomizations for A) simple and B) strict definitions of observation and victimization. Under the simple definition, observation occurs when a cheater kills at least two others while the player is still in the game and victimization occurs every time a player is killed by a cheater. Under the strict definition, observation occurs when a cheater kills at least five others while the player is still in the game and victimization occurs if a player is killed by a cheater when the player is among the last 30\% of survivors.}
\label{fig:fig4}
\end{figure}

The observed interaction effect could be driven by a large number of individuals with a single instance of observation and victimization or by a small number of individuals with more than one instance. To test H4, we next distinguish between contagion events after a single exposure to a cheater ($n_o \leq 1$, $n_v \leq 1$) and contagion events after two or more exposures ($n_o \geq 2$ and/or $n_v \geq 2$). In this case, if a player was never killed by a cheater but observed 2 cheaters, we count this as one ($n_o \geq 2$, $n_v = 0$) event sequence; if a player was killed by 3 cheaters and observed 4 cheaters, we count this as one $(2+, 2+)$ event sequence. The results show that only the ($n_o \geq 2$, $n_v \geq 2$) contagion event sequences are unlikely to occur under the null hypothesis and thus cannot be explained away with the network structure and temporal patterns (Fig.\ \ref{fig:fig4}). This provides evidence for H4.3 that is consistent for both the simple and strict operationalizations of observation and victimization. The evidence will be statistically significant even if we choose to correct for the three simultaneous tests for ($n_o = 1$, $n_v \geq 2$), ($n_o \geq 2$, $n_v = 1$), and ($n_o \geq 2$, $n_v \geq 2$), in which case the significance threshold will be adjusted downward to $p < 0.016$ under the strictest Bonferroni correction \cite{shaffer95}. In contrast, we cannot reject the null hypotheses for H4.1 and H4.2 as the event sequences ($n_o \geq 2$, $n_v = 0$) and ($n_o = 0$, $n_v \geq 2$) are just as common in the randomized networks as in the empirical data.

We note that the results for H4.1--H4.3 replicate the results for H1--H3: even when we consider multiple exposure, there is lack of support for isolated effects from observation and victimization but evidence for an interaction effect between the two. Further, Figure \ref{fig:fig4} reveals that the probability of the ($n_o = 1$, $n_v = 1$) event sequence clears the 0.05 threshold only in the analysis with the strict definition of observation and victimization. Although our method does not allow to compare the size of effects, one possible interpretation of this result is that the interaction effect is more robust and, thus, likely stronger when individuals encounter cheaters on more than one occasion compared to a single exposure.

\section*{Discussion}

In this study, we set out to find empirical evidence for two different social processes that enable the contagion of antisocial behavior: observation and victimization. Most notably, we searched for this evidence in an actual social setting, outside of the lab. To do this, we made use of large-scale digital records of player interactions in online gaming. Our analytical approach was to count event sequences in which a player who observes and/or is killed by a cheater adopts cheating within a week and then compare the count in the empirical data with the counts we expect in alternative gameplay scenarios, where the observer and/or victim happened to be someone else. Our study identifies social contagion because the unique game setting prevents direct and indirect reciprocity as alternative explanations for the adoption of cheating, while the analytical approach we employ controls for homophily and individual and situational factors as additional drivers for the spread of cheating. 

Although we did not find evidence that observation alone increases the adoption of cheating, nor that victimization alone does so, we found that cheating spreads when both observation and victimization are present. This suggests that third-party influence and generalized reciprocity are distinct mechanisms that, even if weak on their own, can interact positively and trigger behavioral change. Additionally, we found the strongest evidence for the interaction of the two mechanisms when both observation and victimization are repeated; this suggests that cheating is more likely to spread after repeated exposure to cheaters. 

Our work extends existing research on the contagion of social behavior in two different ways. First, we differentiate between the effects of observation and victimization in the spread of antisocial behavior, the latter of which has received little attention in the literature so far. The effects of victimization are driven by generalized reciprocity: the tendency to replicate an antisocial act towards an innocent third party as an act of displaced frustration and aggression. In large networks, generalized reciprocity could cause pay-it-forward cascades of contagious malevolence that magnify the negative effect of disadvantaging or harming a single individual. Such cascades are not possible with direct and reputation-based indirect reciprocity. Generalized reciprocity is evident in the context of targeted antisocial behavior such as theft, physical assault, or verbal abuse but it is also plausible for indiscriminate unethical behavior in competitive and resource-constrained environments. Under competitive pressures, individuals may incorrectly approach situations as zero-sum \cite{meegan10} and perceive non-directed antisocial behavior such as academic cheating, employee theft, or tax evasion as unfair and personally damaging. Our research suggests that generalized reciprocity matters because we observe strong enough contagion effects only for the observers of antisocial behavior who are also victims of it.

Our second contribution is that we distinguish between the effects of observation and victimization, and correspondingly, the mechanisms of third-party influence and generalized reciprocity, using observational data. This is notoriously difficult to accomplish because, on the one hand, selection and influence are confounded in social interaction networks, and on the other, generalized reciprocity can rarely be isolated from direct reciprocity, reputation-based indirect reciprocity, and third-party influence outside the lab. Our unique context, data, and method enable us to overcome these challenges. The computational method we employed allowed us to control for the compositional, temporal, and structural patterns in the data driven by the low incidence of the phenomenon we study, the peculiar nature of interactions, and the confounding effect of unobserved friendship relations. The approach we employed is particularly useful for rare events and large data. Both traditional statistical methods, such as logistic regression, and other advanced approaches for longitudinally observed networks, such as temporal exponential random graph models \cite{lusher13}, stochastic actor-oriented models \cite{snijders10}, relational event models \cite{butts08}, and dynamic network actor models \cite{stadtfeld17,stadtfeld17socsci} would struggle with the amount of data we analyze. 

Nevertheless, we also recognize that our study has certain limitations. One drawback of our approach is that it prevents us from estimating an effect size. Counting contagion event sequences allows us to find evidence for the mechanisms we study but not to quantify their strength. But even without a precise estimate, it is clear that the contagion of cheating is weak in the game we study. The fact that the incidence of cheating is extremely low in our data suggests that the severe punishment the game company imposes on cheaters works. We find evidence that contact with cheaters propagates cheating but this does not seem to occur at a rate that can overwhelm and undermine the system.

The low incidence of cheating in the data also makes us vulnerable to Type I errors, where we incorrectly reject the null hypothesis of non-existing effect. We chose not to correct for multiple testing because our tests were prespecified and theory-led, as well as duplicated for two different operationalizations. The estimated proportions and standard errors we report indicate the precise level of uncertainty in our results and hence, we remain cautious about the overall conclusions of the study. There is a clear need for further empirical exploration of how antisocial behaviour spreads. 

Our study also suffers from limitations related to the data. Our measures may not be precise enough since we had to use a number of assumptions and heuristics to identify who is a cheater, estimate when identified cheaters start cheating, and define when players experience and observe cheaters. In particular, we do not know whether victims of cheaters were fully aware that they were harmed by or observed cheaters because it is sometimes difficult to distinguish between cheaters and very skillful players. Players may fail to observe cheating, especially when the cheaters are distant or use less noticeable cheating tools. However, the fact that our results are relatively robust to the two different operationalizations of observation and victimization we used gives us confidence in the findings. 

It is also possible that there are biases in who the detected cheaters are since, for example, skillful players are more likely to be suspected to cheat, and hence, reported, investigated, and banned. However, such a bias would only strengthen our conclusions. First, undercounting low-skill cheaters ensures a more valid measurement of observation because other players are much more likely to notice the existence of cheaters and tempted to adopt cheating when cheaters perform well. Second, undercounting low-skill victims and/or observers who start cheating means that our results are conservative, and that the contagion is even stronger in reality than what we detect. 

Finally, we note that our findings do not align with previous experimental research, which found an effect for victimization but no positive interaction between observation and victimization \cite{tsvetkova15}. However, this discrepancy could be partially explained with the strength of the effects. It is possible that we cannot detect social contagion for those who only observe and those who only experience because the effects are simply too weak to detect in our data. We also note that our tests are highly conservative---we use a very restrictive null model in order to account for all possible sources of confounding and bias. Thus, at the very least, our results imply that social contagion is strongest and most detectable for those who both observe and experience. 

Despite these limitations, our unique social context and data allowed us to study population dynamics in ways that were impossible until recently. Although multiplayer online games are not identical to, nor even representative of, offline interactions, they nevertheless constitute real social systems in which real people bring in their inherent desires to cooperate, compete, gain status, or obtain social approval. Thus, online gaming systems represent a useful research tool to investigate fundamental social mechanisms and emergent social phenomena such as inequality, collective action, or social contagion \cite{szell10,golder14,nishi15}. Studying the contagion of antisocial behavior in an actual social setting allows us to contextualize the mechanisms and their impact, as well as think more concretely about possible strategies to fight the spread of bad influence. Our study brings attention to the social causes of cheating, in addition to any psychological or cognitive sources. Since individuals' interactions with cheaters affect the spread and prevalence of cheating, interventions should also target interactions, not just individuals. But as always, generalizations to other social contexts, whether on- or offline, should be made with caution. People in different social settings may have different attitudes or susceptibilities to peer influence and antisocial behavior. In addition, antisocial behavior varies in the extent to which it is personally beneficial or damaging to others. Both of these characteristics could affect the contagion process. 

Rather than providing a definitive answer to the problem of the contagion of antisocial behavior, we hope that our work inspires further research. There is need both for replications in the lab and generalizations to other social contexts and settings. For instance, our study can easily be repeated to investigate the spread of offensive language in online forums. In cases where detailed data on interactions are not readily available, instead of testing for the mechanisms directly, one can test for their macro-level implications. For example, the finding that victimization positively interacts with observation would imply that antisocial behavior will be more common when it is directly and personally damaging. For instance, academic cheating is considered anti-social behavior but its negative effects are usually diffused. However, when students are graded ``on the curve'', a cheater's unfair advantage directly affects everyone else's grades. Thus, we would expect that academic cheating would be more widespread in schools or courses that grade ``on the curve'' than those that do not and this hypothesis can be tested with observational studies and field experiments. 

In terms of practical applications, our findings point to a number of ways to regulate cheating and antisocial behavior in online games specifically, but potentially also in other social environments. One strategy would be to micro-target individuals who both experience and observe antisocial behavior since they may be at a high risk to adopt it. To take competitive sports as an example, athletes who have in the past narrowly lost to a newly uncovered cheater may be more susceptible to doping or cheating. A complementary strategy would be to control the interactions of antisocial individuals and information about their activity in order to minimize the possibility that any individual would both experience and observe multiple instances of antisocial behavior. Finally, situations and settings that increase the personal costs from others' antisocial behavior should be avoided. Thus, for example, schools and organizations could reduce the spread of cheating by abolishing practices such as grading students ``on the curve'' or evaluating employees with ``forced ranking'', since, with them, one person's cheating has direct effect on another person's performance.

\section*{Acknowledgements}

The authors acknowledge the generous support of the Volkswagen Foundation under Grant Ref. 92 173. The funders had no role in the study design, data collection and analysis, decision to publish, or preparation of the manuscript.

\bibliographystyle{naturemag}
\bibliography{cheating}

\renewcommand{\thefigure}{S\arabic{figure}}
\renewcommand{\thetable}{S\arabic{table}}

\clearpage

\begin{center}

\textbf{\Large{Supplementary Materials for:}} 
\vspace{0.25in}

\LARGE{Cheating in online gaming spreads through observation and victimization}

\vspace{0.25in}
\large{Ji Eun Kim and Milena Tsvetkova}
\end{center}

 \vspace{0.25in}
 
\begin{figure}[h]
\centering
\includegraphics[width=\textwidth]{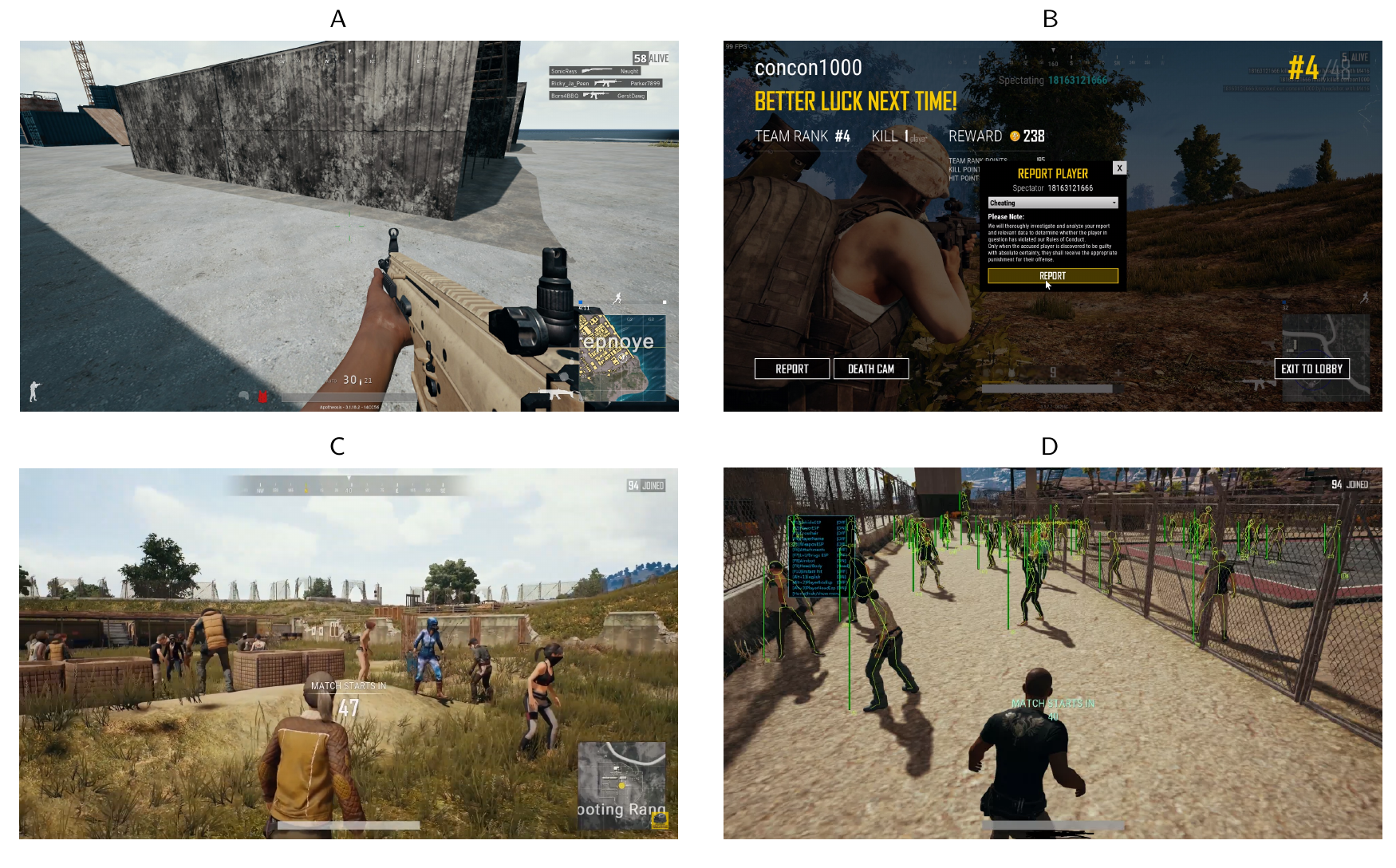}
\caption{A) Screenshot from a match of PlayerUnknown's Battlegrounds. Kill feeds (upper right corner) display killers and their victims in real time. Image source: \url{https://i.redd.it/tef4q5rbfwxz.jpg}. B) Screenshot of the in-game report system. Players can report suspected cheaters after watching their ``death cams.'' The death cam allows players to see a short video showing how they died. Although not shown here, upon dying, players also have the option to continue watching the game from their killer's perspective. Image source: \url{https://www.youtube.com/watch?v=907cWYjP-gU}. C) Screenshot from the waiting time before a match starts showing how a regular player views others. Image source: \url{https://www.youtube.com/watch?v=lHhcGiQk4W0&t=5s}. D) Screenshot from the same game stage but for a player who is using a cheating tool known as ``ESP hack.'' With this tool, the cheater can see other players through walls and locate the position of nearby opponents. Image source: \url{https://www.gamesradar.com/pubg-cheats-explained/}.} 
\label{fig:figs1}
\end{figure}

\begin{figure}[h]
\centering
\includegraphics[scale=1]{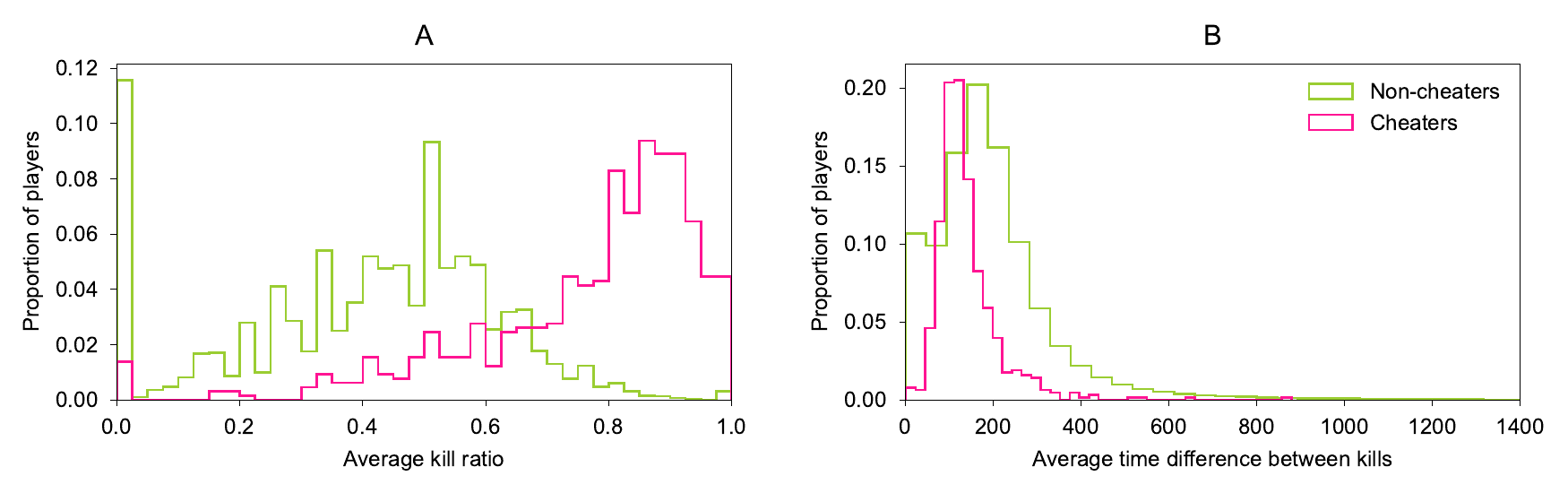}
\caption{A) Histogram of the average kill ratio for cheating and non-cheating players. B) Histogram of the average time difference between consecutive kills for cheating and non-cheating players (outliers are excluded).}
\label{fig:figs2}
\end{figure}

\begin{figure}[h]
\centering
\includegraphics[scale=1]{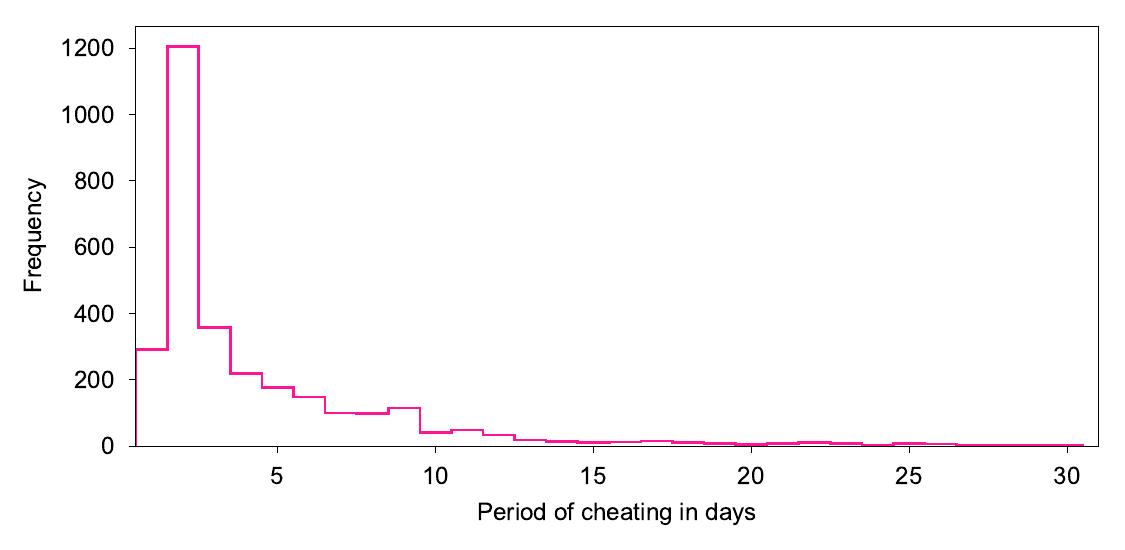}
\caption{Histogram of the period of cheating for the 2,980 cheaters with full information on performance.}
\label{fig:figs3}
\end{figure}

\begin{figure}[h]
\centering
\includegraphics[scale=1]{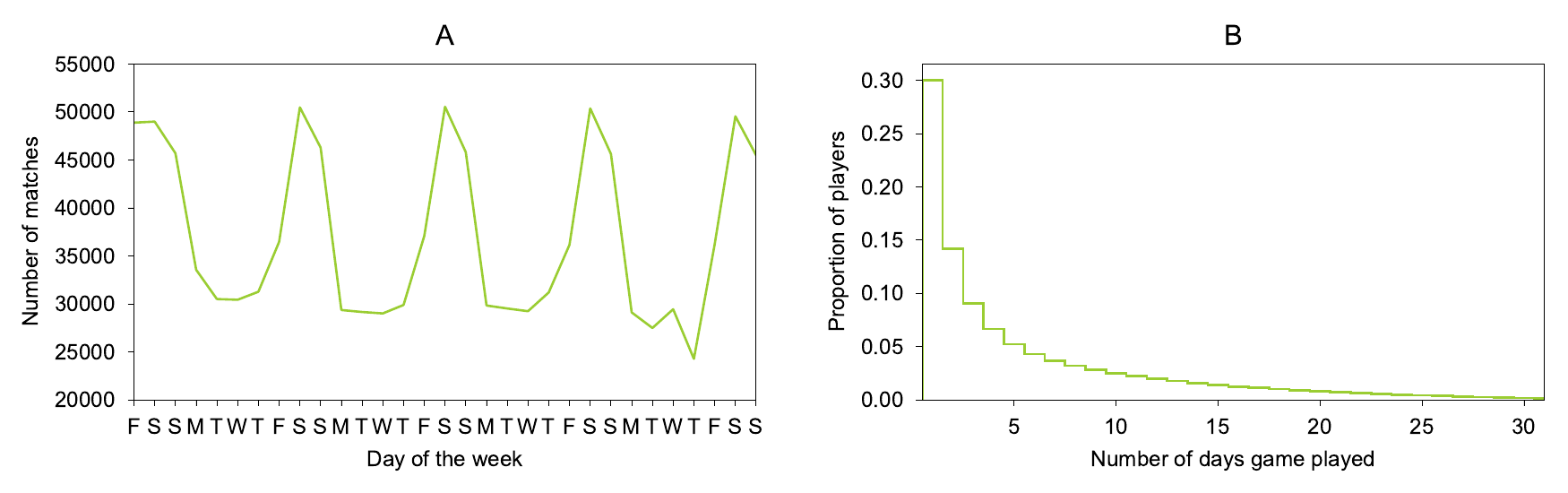}
\caption{A) Number of matches played by day for the period March 1--31, 2019. The sudden dip in the last trough is due to the game server being down between 9:30 am and 4:30 pm on March 28th as part of routine maintenance. B) Histogram of the number of days players accessed the game during the month-long data collection period.}
\label{fig:figs4}
\end{figure}

\begin{figure}[h]
\centering
\includegraphics[scale=1]{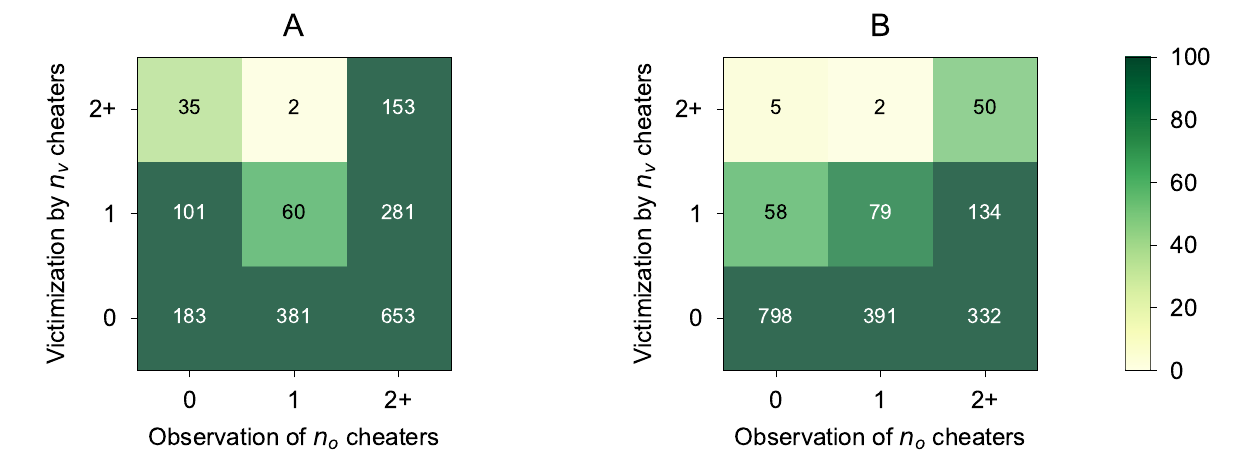}
\caption{The frequency of candidate contagion events in the empirical data for A) simple and B) strict definitions of observation and victimization. Under the simple definition, observation occurs when a cheater kills at least two others while the player is still in the game and victimization occurs every time a player is killed by a cheater. Under the strict definition, observation occurs when a cheater kills at least five others while the player is still in the game and victimization occurs if a player is killed by a cheater when the player is among the last 30\% of survivors.}
\label{fig:figs5}
\end{figure}

\begin{figure}[h]
\centering
\includegraphics[scale=1]{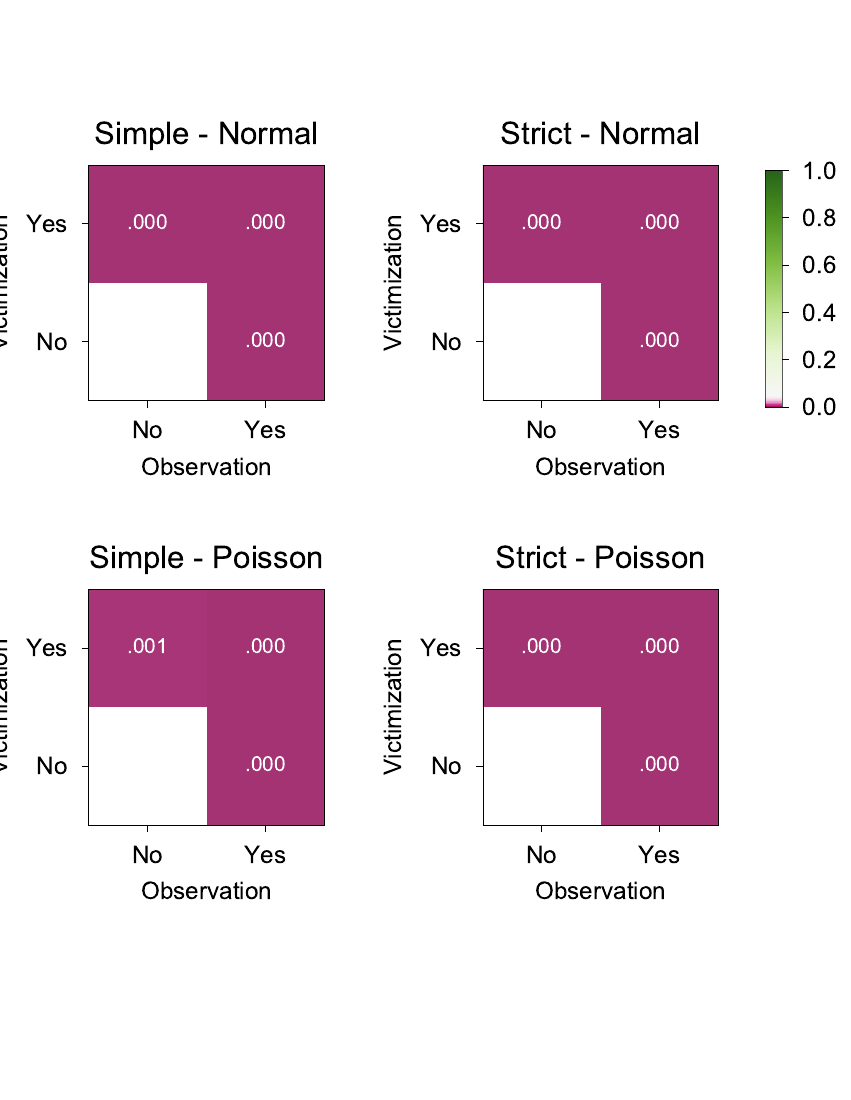}
\caption{$\chi^2$ goodness-of-fit tests for whether the motif counts from the 1,000 network randomizations follow a normal distribution with the same mean and standard deviation. Cell numbers and colors show the $p$-value for testing the null hypothesis that the two distributions are identical; a small $p$-value ($p < 0.05$) means that the null hypothesis can be rejected and the two distributions are sufficiently different.}
\label{fig:figs6}
\end{figure}

\begin{figure}[h]
\centering
\includegraphics[scale=1]{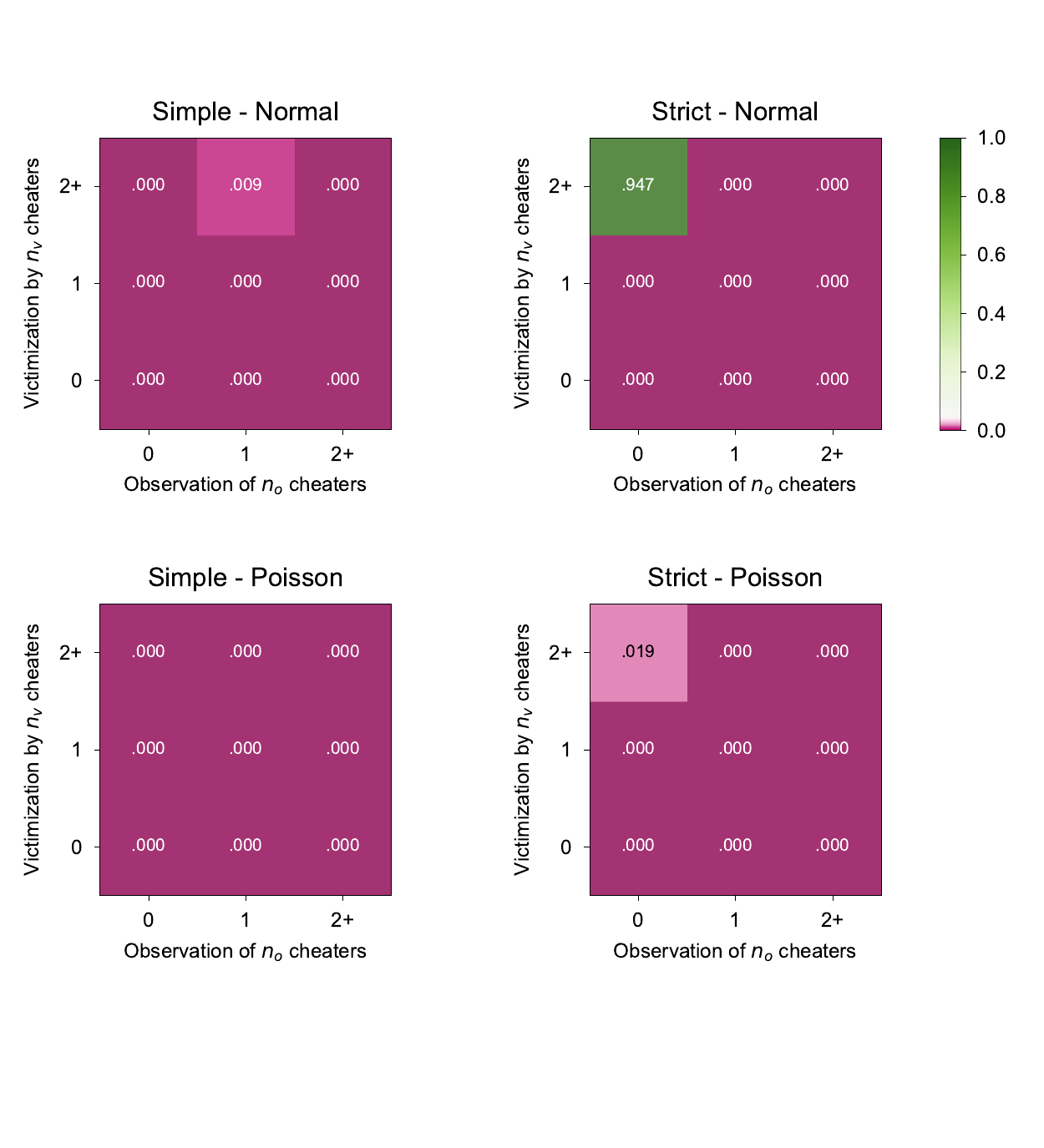}
\caption{$\chi^2$ goodness-of-fit tests for whether the motif counts from the 1,000 network randomizations follow a Poisson distribution with the same mean rate. Cell numbers and colors show the $p$-value for testing the null hypothesis that the two distributions are identical; a small $p$-value ($p < 0.05$) means that the null hypothesis can be rejected and the two distributions are sufficiently different.}
\label{fig:figs7}
\end{figure}

\clearpage

\begin{table} \footnotesize \centering
\caption{Number of matches and number of matches with cheaters by game mode.}
\begin{tabular}{cccccccc}
\hline
\hline
 \textbf{Game mode} &  \textbf{Num.\ matches} & \multicolumn{6}{c}{\textbf{Num.\ cheaters}} \\
& & \textbf{1} & \textbf{2} & \textbf{3} & \textbf{4} & \textbf{5} & \textbf{6} \\
\hline
Solo & 124,421 (10.8\%)	& 10,265	& 951	& 77		& 3		& 0	& 0 \\
Duo & 359,870 (31.4\%)	& 19,195 	& 1,730	& 135	& 11		& 0	& 0 \\
Squad & 662,650 (57.8\%) & 64,095	& 9,079	& 1,355	& 212	& 27	& 4 \\
\hline
Total & 1,146,941 (100.0\%) & 93,555 & 11,760	& 1,567	& 226	& 27	& 4 \\
\hline
\end{tabular}
\end{table}

\begin{table} \footnotesize \centering
\caption{Rate of success for cheating solo players and for teams with cheating teammates.}
\begin{tabular}{lccc}
\hline
\hline
\textbf{Num.\ cheaters} & \textbf{Num.\ players} & \textbf{Prop.\ who win} & \textbf{Prop.\ in top 30\%} \\
\hline
\multicolumn{4}{l}{\textbf{Solo matches}} \\
0 & 1,049,113 & 0.01 & 0.28 \\
1 & 12,410 & 0.13 & 0.42 \\
\multicolumn{4}{l}{\textbf{Duo and squad matches}} \\
0 (single-player team) & 143,791 & 0.00 & 0.19 \\
1 (single-player team) & 4,889 & 0.08 & 0.30 \\
0  & 2,807,526 & 0.03 & 0.38 \\
1  & 96,747 & 0.15 & 0.55 \\
2  & 4,094 & 0.26 & 0.60 \\
3  & 187 & 0.39 & 0.69 \\
4  & 11 & 0.82 & 0.91 \\
\hline
\multicolumn{4}{p{12cm}}{\textit{Note:} The statistics are calculated over the 107,139 matches with at least one cheater. 
} \\ 
\end{tabular}
\end{table}

\end{document}